\documentclass[fleqn,10pt]{wlscirep}
\usepackage[utf8]{inputenc}
\usepackage[T1]{fontenc}
\usepackage{lineno}
\usepackage{tcolorbox}
\usepackage{multirow}
\usepackage{multicol}
\usepackage[page]{appendix} 
\usepackage{ltablex} 
\usepackage{dirtree}
\usepackage{graphicx}

\title{Endomapper dataset of complete calibrated endoscopy procedures}

\author[1]{Pablo Azagra}
\author[2,3,4,5]{Carlos Sostres}
\author[2,3,4,5]{Ángel Ferrández}
\author[1]{Luis Riazuelo}
\author[1]{Clara Tomasini}
\author[1]{O. León Barbed}
\author[1]{Javier Morlana}
\author[1]{David Recasens}
\author[1]{Víctor M. Batlle}
\author[1]{Juan J. Gómez-Rodríguez}
\author[1]{Richard Elvira}
\author[2]{Julia López}
\author[1]{Cristina Oriol}
\author[1]{Javier Civera}
\author[1]{Juan D. Tardós}
\author[1]{Ana C. Murillo}
\author[2,3,4,5]{Angel Lanas}
\author[1]{José M.M. Montiel}
\affil[1]{Instituto de Investigación en Ingeniería de Aragón (I3A). Universidad de Zaragoza.}
\affil[2]{Digestive Disease Service. Hospital Clínico Universitario Lozano Blesa, Zaragoza, Spain}
\affil[3]{Department of Medicine. Universidad de Zaragoza, Spain}
\affil[4]{Instituto de Investigación Sanitaria Aragón (IIS Aragón). Zaragoza, Spain.}
\affil[5]{Centro de Investigación Biomédica en Red, Enfermedades Hepáticas y Digestivas (CIBEREHD)}

\begin{abstract}

Computer-assisted systems are becoming broadly used in medicine. In endoscopy, most research focuses on the automatic detection of polyps or other pathologies, but localization and navigation of the endoscope are completely performed manually by physicians. To broaden this research and bring spatial Artificial Intelligence to endoscopies, data from complete procedures is needed. This paper introduces the Endomapper dataset, the first collection of complete endoscopy sequences acquired during regular medical practice, making secondary use of medical data. Its main purpose is to facilitate the development and evaluation of Visual Simultaneous Localization and Mapping (VSLAM) methods in real endoscopy data. The dataset contains more than 24 hours of video. It is the first endoscopic dataset that includes endoscope calibration as well as the original calibration videos. Meta-data and annotations associated with the dataset vary from the anatomical landmarks, procedure labeling, segmentations, reconstructions, simulated sequences with ground truth and same patient procedures. The software used in this paper is publicly available. 

\end{abstract}
\begin{document}

\flushbottom
The Version of Record of this contribution is published in Scientific Data journal, and is available online at: \url{https://doi.org/10.1038/s41597-023-02564-7}
\maketitle

\thispagestyle{empty}

\section*{Background \& summary}

Endoscopes traversing body cavities are routine. However, their potential for navigation assistance or device autonomy remains mostly locked. In order to unlock it, computer-assisted endoscopes would require spatial AI (Artificial Intelligence) capabilities, i.e., being able to estimate a map of the regions that are traversed, along with the endoscope localization within such map. This capability is known in the robotics literature with the acronym VSLAM (Simultaneous Localization and Mapping from Visual sensors). Spatial AI and VSLAM will augment endoscopies with novel features, including augmented reality insertions, detection of blind zones, polyp measurements or guidance to polyps found in previous explorations. In the long term, VSLAM will also support utterly new robotized autonomous procedures. For our purposes in this paper, the goal of VSLAM is to build a per-patient map, in real-time, during endoscope insertion in a first procedure. This map will be exploited and perfected during the withdrawal of such first procedure, and in any other future one. 
 
 \begin{figure}[!ht]
\centering
\includegraphics[width=0.90\textwidth]{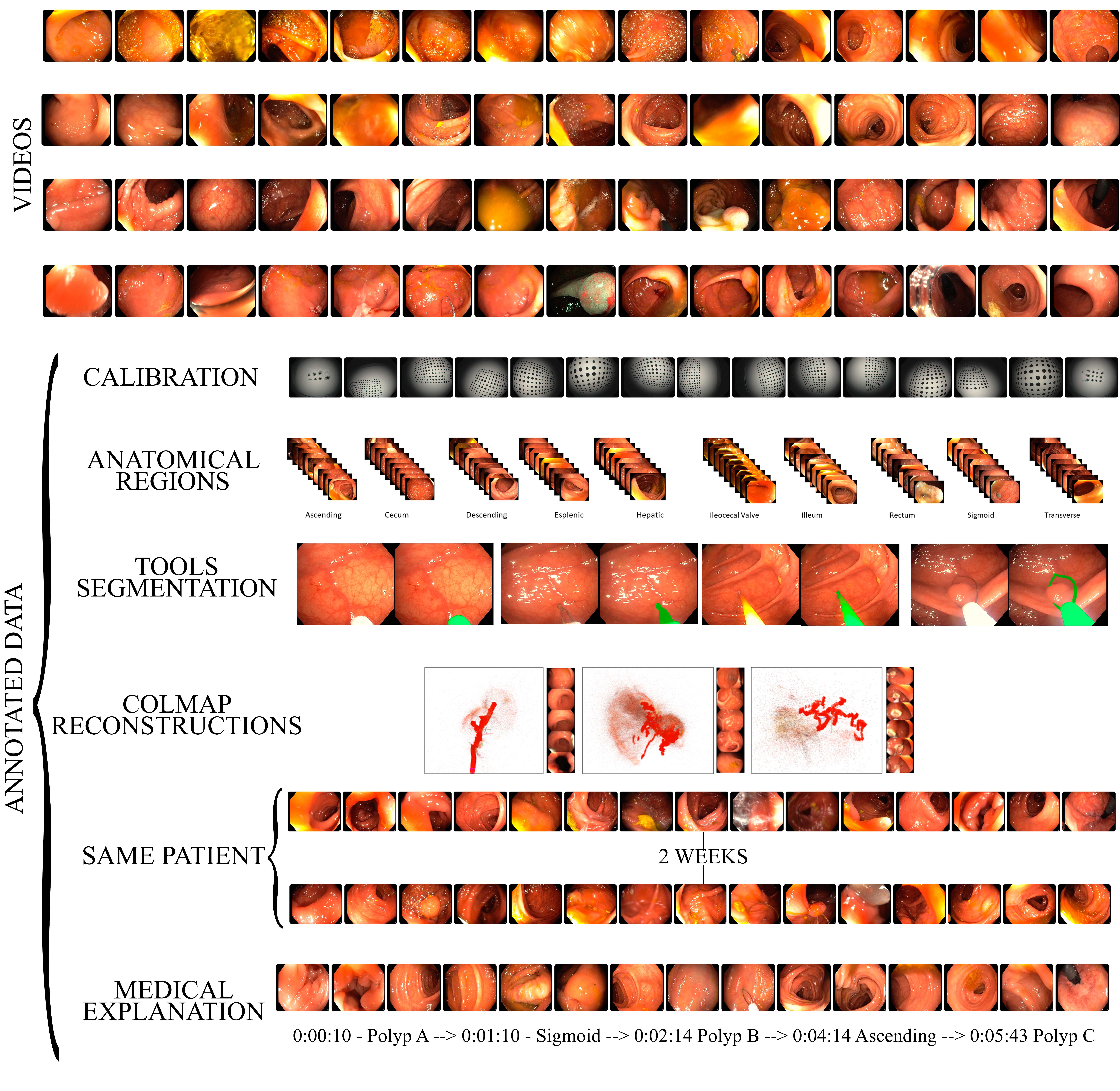}
\caption{Overview of the Endomapper Dataset.}
\label{fig:main}
\end{figure}
 
There are mature methods for out of the body VSLAM \cite{campos2021orb,engel2017direct}. However, bringing them to endoscopy implies overcoming new barriers. The light source is co-located with the endoscopes, and hence is moving and is close to the body surfaces. The body surfaces have poor texture and abundant reflections due to fluids. The scene geometry includes a prevalent deformation. The video combines slow observation of areas of interest, with fast motions and long occlusions of the endoscope lenses.

 Our contribution in this paper is the Endomapper dataset~\cite{EndomapperDataset}, which makes available, for the first time, \textbf{96 high quality calibrated recordings of complete routine endoscopies} (Figure~\ref{fig:main}), making secondary use of medical data, i. e., just recording standard procedures that were going to be performed in any case, without any modification. Compared to ad-hoc recordings, secondary-use ones show realistic features and hence contain the actual challenges VSLAM will face in routine practice. 
 
 No other public dataset offers a comparable volume of fully calibrated endoscopies in HD (see Table~\ref{tab:related}). Heilderberg~\cite{maier2021heidelberg} is a very interesting dataset that contains images of the colon from laparoscopic procedures. However, this view of the colon is not compatible with our goal of building 3D reconstructions of the interior of the gastrointestinal tract. CVC\-ClinicDB, GIANA and Kvasir focus on polyps detection since they are often used to benchmark CAD (Computer Aided Diagnosis) systems. Other datasets focus in segmentation of elements of interest, such as tools in instrument-kvasir or polyps in Kvasir-seg. However, they only provide sparse image sets or short videos (less than 30 seconds). More similar to ours, Colon10k provides images from short sequences for place recognition and reconstructions. In contrast, we offer hours of real calibrated video, corresponding to the complete procedures. Apart from these real imaging datasets, it is worth mentioning works that create simulated data of the colon, either using 3D models, like Rau et al.~\cite{rau2022bimodal}, Bobrow et al.~\cite{bobrow2022colonoscopy} and Incetan et al.~\cite{incetan2021vr}, or a realistic phantom, like Ozyoruk et al.~\cite{ozyoruk2020endoslam}. Similar to these works, we also include a few simulated sequences to harness the extra information of these scenes as a means to evaluate the methods developed. 
 In particular, due to the monocular nature of the dataset, no ground truth geometry is available for quantitative evaluation. To address this issue, we include  photorealistic sequences from a simulated colon, with ground truth geometry for the deforming scene and endoscopy trajectory.

The Endomapper dataset includes colonoscopies, gastroscopies, and calibration videos along with geometric and photometric calibration parameters. More than half of the sequences are screening colonoscopies, for which the standard procedure implies a thorough and slow exploration, being close to typical operation modes for VSLAM and that can serve as a bridge to more challenging sequences. To research map reuse and recognition in a second exploration, colonoscopies corresponding to the same patient but separated in time by several weeks are also included in the data.

\begin{table}
\footnotesize

\begin{tabular}{|l|l|l|l|l|}
\hline
\textbf{Dataset} & \textbf{Purpose} & \textbf{Type of Data} & \textbf{Size of Dataset} & \textbf{Availability} \\ \hline
CVC-ClinicDB~\cite{CVC_clinic} & Polyps segmentation & Images & 612 images & Open Academic \\ \hline
Endoscopic artifact detection~\cite{EndoChallenge} & Artifact Detection & Images & 5,138 images & Open Academic \\ \hline
GIANA 2021~\cite{Bernal2021} & \begin{tabular}[c]{@{}l@{}}Polyp detection, segmentation\\ and classification\end{tabular} & Short Videos and Images & 38 videos and 3000 images & By request \\ \hline
Kvasir~\cite{Kvasir} & \begin{tabular}[c]{@{}l@{}}Anatomical landmarks, \\ Pathological findings, \\ Therapeutic interventions and\\ Quality of mucosal views\end{tabular} & Images & 4000 images & Open Academic \\ \hline
Kvasir-Seg~\cite{jha2020kvasir} & Polpy segmentation & Images & 8000 images & Open Academic \\ \hline
Nerthus~\cite{Nerthus} & Bowel preparation & Short Videos & 21 videos ( 5525 frames) & Open Academic \\ \hline
HyperKvasir~\cite{borgli2020hyperkvasir} & \begin{tabular}[c]{@{}l@{}}Anatomical landmarks,\\ Pathological findings, \\ Therapeutic interventions and \\ Quality of mucosal views\end{tabular} & \begin{tabular}[c]{@{}l@{}} Images \\ \& Short Videos \end{tabular} & \begin{tabular}[c]{@{}l@{}} 110079 images (10662 labeled) \\ \& 374 short videos \end{tabular} & Open Academic \\ \hline
Colon10k~\cite{Ruibin2021} & Place recognition & Images & 10126 images & Open Academic \\ \hline
Instrument-kvasir~\cite{jha2021kvasir} & Tool segmentation & Images & 590 images & Open Academic \\ \hline
Endomapper (ours) & \begin{tabular}[c]{@{}l@{}}VSLAM \end{tabular} & \textbf{Complete  real endoscopies} & \textbf{96 videos ($\sim$24 hours)} & By Request \\ \hline
\end{tabular}
\caption{Overview of existing datasets of endoscopies in the gastrointestinal tract.}
\label{tab:related}
\end{table}

 Regarding metadata, some endoscopies include a description of the procedure made by the endoscopist, in the form of text footage. The text describes the anatomical regions traversed, re-explorations of the same region, the performed interventions or the tools used. This footage indexes the videos to identify interesting sections for VSLAM.
 
  Building on our dataset, the community can provide derived or metadata results to support subsequent research. Some examples of these derived data are included in the dataset: 1) anatomical regions segmentation, at frame level, performed by a doctor after visualizing the video; 2) tools segmentation in selected video sections, which can boost the tool segmentation performance in the specific endoscopy domain; 3) Structure from Motion (SfM) reconstructions using COLMAP \cite{schoenberger2016sfm}, which provides up-to-scale 6 DoF endoscope trajectory and 3D models for the video segments corresponding to smooth explorations of non-deforming scenes. The SfM output has proven valid to supervise learning tasks such as image matching \cite{dusmanu2019d2} or image retrieval \cite{radenovic2018fine}.
   
 Endomapper offers a sweet point of challenge, including easy video segments where state-of-the-art algorithms perform reasonably. However, all these methods also fail at some point, signaling what are the research challenges to face. We believe that the dataset will spur research that identifies challenges and foster progress of VSLAM in gastrointestinal environments.

Finally, we have made publicly available 7 software repositories corresponding to photometric and geometric calibration from calibration videos, simulated colon sequences generation and technical validation. Section \emph{Code Availability} contains the details of these repositories containing the software, including installation and usage instructions.

\section*{Methods}

The methodology used to create the dataset is explained in this section. First, we present a description of the recording procedure for the sequences in the dataset, including the description of the capture system and the type of recordings. Then, we describe the calibration procedure and the methodology used in both geometric and photometric calibration. Finally, we also briefly summarize the methods used to create each type of meta-data.

\subsection*{Recording endoscopies procedure}\label{grabbingWorkstation}

The acquisition of the sequences in the dataset was performed in the Hospital Clinico Universitario Lozano Blesa, in Zaragoza (Spain), using an Olympus
EVIS EXERA III CV-190 video processor, EVIS EXERA III CLV-190 light source, and EVIS EXERA III CF-H190 colonoscope or EVIS EXERA III GIF-H190 gastroscope. The acquisition system is composed of a computer and a data acquisition card connected to the endoscopy tower via a Digital Visual Interface (DVI). Two different acquisition cards have been used: Epiphan Video DVI2USB 3.0 and Magewell Pro Capture DVI. The videos were recorded at 1440×1080 at 40fps and 24RGBbits (Epiphan) or 1440×1080 at 50fps and 24RGBbits (Magewell). The output image given by the endoscopy tower is cropped to remove personal information. The videos were manually edited to remove any frame recorded when the camera was out of the body of the patient. During the span of the project, the recordings were done one day of the week and synchronized with the medical staff involved. The patients were not selected based on their symptoms or pathology, we followed without interference the hospital's schedule which was mostly focused on the colorectal cancer screening program. The Endomapper technical staff was present in all the recording sessions to secure the quality of the acquisitions, but without interfering with the medical procedure.

\subsection*{Use of human participants}
The recordings were made under the ethical approval of the CEICA Ethics Committee ({\emph{Comité de Ética de la Investigación de la Comunidad Autónoma de Aragón (CEICA)}, meetings 04/03/2020 acta 05/2020, 23/09/2020 acta 18/2020, 20/04/2022 acta 08/2022 and 16/11/2022 acta 20/2022). Informed consent was obtained from all subjects. According to this approval, the collection can be publicly accessed under certain conditions (see Section \emph{Usage Notes}).}

\subsection*{Calibration}\label{calibration}

\begin{figure}[htb]
\centering
\includegraphics[width=\textwidth]{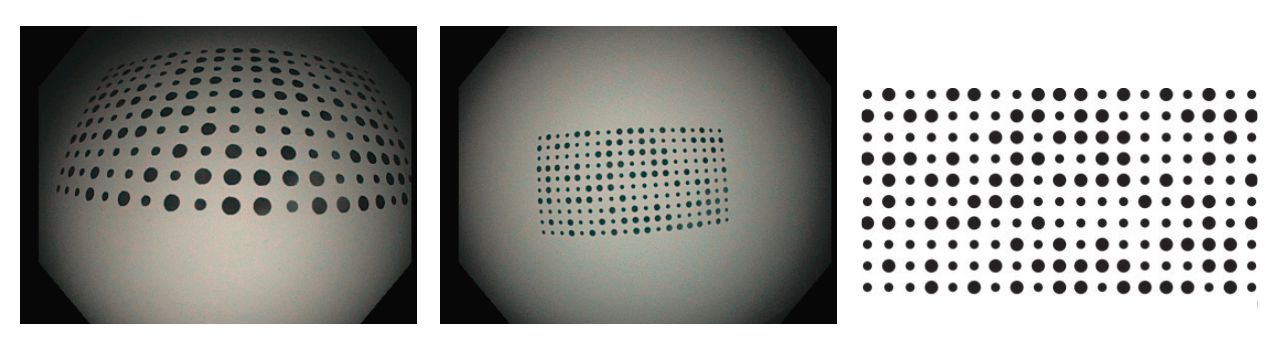}
\caption{Two examples of calibration images (left, middle). The calibration pattern (right)}
\label{im:calib_staff}
\end{figure}

The dataset uses 10 different colonoscopes and 8 different gastroscopes. The calibration sequences for all the colonoscopes and gastroscopes where acquired in a single session using a Lambertian pattern (obtained from \url{calib.io}). Figure \ref{im:calib_staff} shows two frames of the calibration videos imaging the calibration pattern. The Lambertian pattern corresponds to an array of circles from the Vicalib~\cite{Vicalib} library. The physical size of the pattern used is $5,61\times9,82$\,cm.

\subsubsection*{Geometric calibration}

The calibration videos are processed by Vicalib~\cite{Vicalib} to obtain the endoscope intrinsic parameters according to the Kannala \& Brandt model~\cite{kannala2006generic,usenko2018double}.  The calibration defines eight intrinsic parameters, four projective parameters (in pixels)  $f_x, f_y, C_x, C_y$ and four distortion coefficients  $k_1,k_2, k_3,k_4$. We process 1 out of 20 frames and outlier matches are removed. 
Next, the projection model yielding the projection in pixels $\mathbf{u}=\left(u,v\right)$, for a 3D point with coordinates $\mathbf{X}=\left(x,y,z\right)$ with respect to the camera frame is described as:
\begin{eqnarray}
\centering
u&=&f_x x_d + C_x, \;\;\;\;\;\;\;\;\; x_d=r_d\frac{x}{r}\\
v&=&f_y y_d + C_y, \;\;\;\;\;\;\;\;\; y_d=r_d\frac{y}{r}
\end{eqnarray}
where $r_d=\theta\left(1 + k_1 \theta^2+k_2  \theta^4+k_3 \theta^6+k_4 \theta^8\right)$ is the distorted radius, $r = \sqrt{x^2+y^2}$ is the undistorted radius and  $\theta=\arctan2\left(r,z\right)$ is the angle between the incoming ray and the optical axis.

\subsubsection*{Photometric calibration}
The light source and camera of the endoscope are calibrated to obtain a model able to reproduce the photometry of the recordings. In the endoscope, the distances between the light sources and the camera are small and mostly symmetrical. Following  Modrzejewski et al.~\cite{modrzejewski2020light}, we assume that these sources can be modelled as a single virtual light and adopt the Spot Light Source model (SLS), which was shown to offer a good compromise between complexity and accuracy. In addition, the light spread function and the camera vignetting are jointly estimated assuming radial symmetry. With this model, the light radiance going from the endoscope to a 3D surface point $\mathbf{X}$ is
\begin{equation}
    \sigma_\text{SLS}(\mathbf{X},\, \mathbf{P})
    =
    \sigma_0 \; R(\mu,\, \mathbf{D},\, \mathbf{X},\, \mathbf{P})\; S(\mathbf{X},\, \mathbf{P})\; L(\mathbf{X},\, \mathbf{P}),
\end{equation}
\noindent where $\mathbf{P}$ is the light center, $\sigma_0$, is the light's intensity value and $\mathbf{D}$ is the principal direction in which light propagates with a spreading factor $\mu$, that modules the radial attenuation $R$. As the light traverses the scene, its radiance decreases as a function of the distance travelled $d = ||\mathbf{X} - \mathbf{P}||$, following an inverse-square law $S(\mathbf{X},\, \mathbf{P}) = 1/d^2$. Finally, $L(\mathbf{X},\, \mathbf{P})$ is the unit vector of the light's outgoing direction. The corresponding intensity value $\mathcal{I}(\mathbf{X})$ on the image is:

\begin{equation}
\label{eq:photometric-model}
    \mathcal{I} \left( \mathbf{X} \right)
    =
    \left(
  \lvert \sigma_\text{SLS}(\mathbf{X},\, \mathbf{P}) \rvert \ 
  f_r(\boldsymbol{\omega}_i, \boldsymbol{\omega}_r) \ 
  \cos \theta \
  g_t
    \right)^{1/\gamma},
\end{equation}
where a bidirectional reflectance distribution function (BRDF)  $f_r(\boldsymbol{\omega}_i, \boldsymbol{\omega}_r)$ defines how light is reflected from the surface to the camera. The projection of the light beam on the geometry introduces a cosine term of the angle $\theta$ between the incoming light ray $\boldsymbol{\omega}_i$ and the surface normal. Finally, the endoscope applies an automatic gain $g_t$, that can vary at every $t$-th time instant, and a gamma curve ($\gamma = 2.2$) to improve the perceived dynamic range of the image.

\begin{figure}
    \centering
	\includegraphics[width=\textwidth]{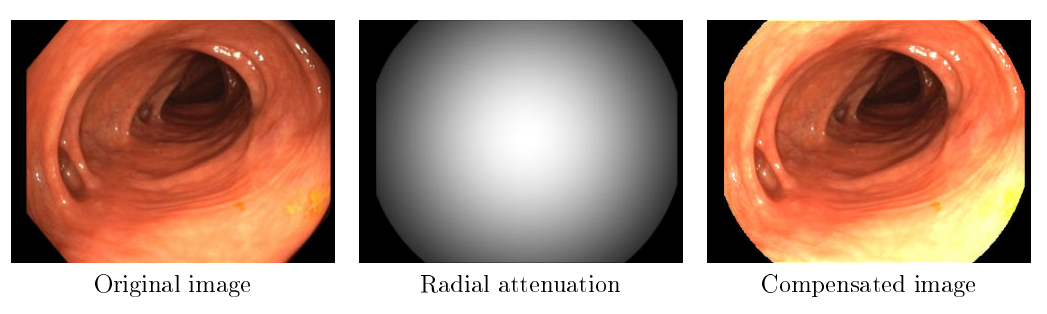}
    \caption{Example of photometric calibration results.}
    \label{im:calib_photo}
\end{figure}

The parameters of this model are estimated by optimising a photometric loss on the white areas of the Vicalib pattern (Figure~\ref{im:calib_staff}). The results of the calibration provide a 2D weighting of the photometric effects caused by the vignetting and the light spread function, that can be used to compensate them (Figure~\ref{im:calib_photo}).

\subsection*{Simulated colon}
The VR-Caps~\cite{incetan2021vr} simulator is used to generate photorealistic synthetic image sequences of a 3D colon model obtained from a Computed Tomography. Since this is a simulation, we have full access to scene configuration: camera calibration, deformations, trajectory and illumination, hence to the ground truth geometry, camera pose and 3D deforming scene. For the same endoscope trajectory, we generated different sequences with more aggressive deformations to allow ablative studies with respect to the deformation magnitude. Deformations applied are described by the next equation:
\begin{equation}
    V_y^t = V_y^0 + A \sin(\omega t + V_x^0 + V_y^0 + V_z^0),
\end{equation}
where $V_x^0$, $V_y^0$ and $V_z^0$ are the coordinates of the surface point at rest. We can control the magnitude and velocity of the deformations according to the parameters $A$ and $\omega$ respectively, which corresponds to the maximum excursion and velocity of the deformations respectively. We also modified the colon texture to increase its contrast. The code to create these simulated sequences is available in the repository \textit{EM\_Dataset-Simulations} (see Section \textit{Code availability},~\url{https://github.com/endomapper/EM_Dataset-Simulations} ).

\subsection*{Meta-data}

For a set of selected recordings, we provide several types of meta-data useful for plenty of potential research lines, but in particular for VSLAM. This subsection presents a description of the meta-data and the annotation methodologies.

\subsubsection*{Text footage}

The endoscopist performing each procedure provided a description of it, that was registered during the exploration. It includes the anatomical regions traversed, the interventions, the medical findings such as polyp approximated size, the tools used, or the sections with NBI (Narrow-Band Imaging) illumination. This description is made available as text footage synchronized with the corresponding videos.
This metadata can be useful, for example, to identify the sections of the video that are more promising for VSLAM, such as the re-observations of the same region or interactions with tools of known size.

\subsubsection*{Anatomical regions}
\begin{figure}[h!]
\centering
\includegraphics[width=0.90\textwidth]{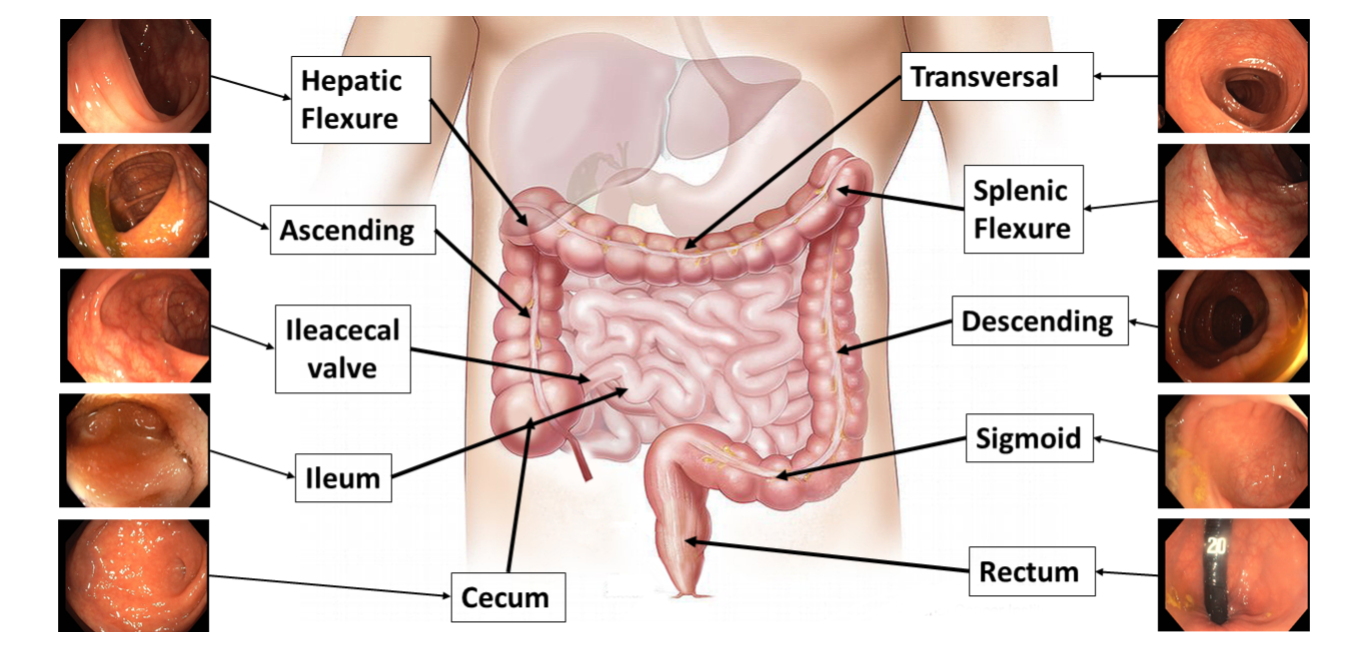}
\caption{Illustration of the anatomical regions labeled.}
\label{fig:places}
\end{figure}

Anatomical section recognition is useful to create topological maps of the colon. These maps can be used to create smaller reconstructions with less probability of errors. Some colonoscopy procedures were annotated by the medical staff of the project after the recording. Multiple careful visualizations were necessary to delimit the ten anatomical regions, that are shown in Figure~\ref{fig:places}.

\subsubsection*{Tools segmentation}
Tool segmentation is one of the challenges for spatial AI in colonoscopies. Since they occlude the view and cause failures in other algorithms, many works in the literature mask them out. Tools were manually segmented using Odin CAT tool\cite{odinCat}, which allows to maintain a mask between frames, giving a more robust annotation.

\subsubsection*{COLMAP 3D reconstruction}
Traditional SIFT-based rigid SfM algorithms are able to produce partial reconstructions from colonoscopy videos. We include some examples of the output of COLMAP \cite{schoenberger2016mvs,schoenberger2016sfm} processing in our sequences, which provides a first approximation for the up-to-scale camera trajectory and the scene's sparse structure. 
This information can be organized to produce weak supervision in the form of sparse depth maps, local correspondences between frames, image-to-image labels (frames depicting the same place) or relative camera pose transformation between frames. Several computer vision tasks like depth prediction, image matching, image retrieval and visual localization can greatly benefit from this kind of supervision. Megadepth \cite{MegaDepthLi18} is a well-known dataset that uses this SfM procedure to obtain 3D point clouds, similar to ours. It is being extensively used for deep learning supervision \cite{dusmanu2019d2, sarlin2021back, yang2020ur2kid}. Other works employed SfM to identify co-visible frames in the recordings, which has proven to be useful to train CNNs for place recognition in landmark images \cite{radenovic2018fine} and in colonoscopy sequences \cite{morlana2021self,Ruibin2021}. 

For our recordings, we apply exhaustive guided matching between all the images in the sequence to associate frames that are temporally distant. We use our camera calibration and we do not optimize it during the COLMAP bundle adjustment. The minimum triangulation angle is relaxed to 8 degrees during the initialization of the models. The rest of the parameters are left as default.

\subsubsection*{Recordings from the same patient}
One of the main obstacles in colon reconstruction is the consistency between colonoscopies in longitudinal studies. Thanks to the colorectal cancer screening program, colonoscopy pairs from the same patient were registered. This would help to evaluate the lifelong capabilities of the developed VSLAM algorithms.

\section*{Data Records}
\begin{figure}[h]
\centering
 \fbox{\includegraphics[trim=0 0 8cm 0,width=0.8\textwidth]{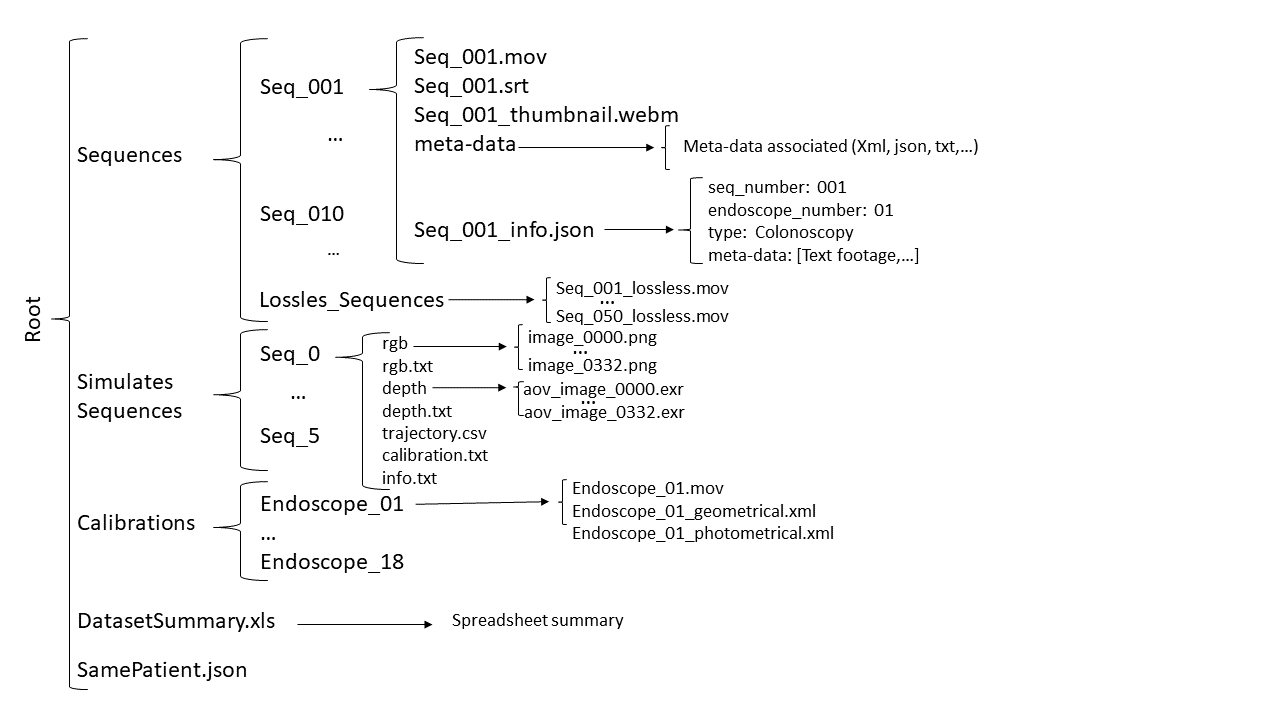} }
\caption{Directory structure of the dataset. 
}\label{fig:structure}
\end{figure}

This section describes the dataset structure and details of the meta-data available. The dataset is available on the Synapse platform\cite{EndomapperDataset} and is subject to access controls (see Section \emph{Usage Notes}). A summary of the dataset structure can be seen in Figure~\ref{fig:structure}. At publication time, there is a total of 96 real sequences and their duration goes from less than ten minutes to more than half an hour. The file {\tt DatasetSummary.xls} in the dataset main folder includes a summary of the acquisition details of each sequence in the dataset.

\subsection*{Video recordings}
Data is stored in the directory {\tt Sequences}. Each procedure has a corresponding directory {\tt Seq\_XXX} ({\tt XXX} is the sequence number) that contains:
\begin{enumerate}
\item The directory {\tt meta-data}, that contains all the meta-data files associated to the sequence. These files are described in the next section.
\item The video {\tt Seq\_XXX.mov}, in which the actual recording is. The video codec is H264~\cite{richardson2004h}, a lossy compression using the profile High 4:4:4 with 4.2 level and a bit rate of 7Mbps.
It offers an optimal size vs. quality trade-off for lossy compression. 
\item The thumbnail version, {\tt Seq\_XXX\_thumbnail.webm}, that contains a compressed version of the recording for easy and quick visualization. This version uses the free codec {\tt libvpx}~\cite{grange2016vp9}, at $320\times240$ resolution.
\item A subtitle file, {\tt Seq\_XXX.srt}, if the video has text footage in the form of text subtitles.
\item The metadata file, {\tt Seq\_XXX\_info.json}, where sequence number, endoscope number and the type of metadata of the procedure is stored. 
\end{enumerate}

Additionally, the folder {\tt Lossless\_sequences} contains the lossless versions of the videos. This format uses codec ffv1 version 3 with a bitrate of 310 Mbps. 

\subsection*{Camera calibration}
All the calibration information is included in the directory {\tt Calibrations}. There is a directory {\tt Endoscope\_XX} ({\tt XX} is the endoscope number) for each endoscope that contains:
\begin{enumerate}
\item The calibration video {\tt Endoscope\_XX.mov}. This version is the lossy H264 version. The lossless version can be found in the lossless folder mentioned before.
\item The Geometric calibration parameters {\tt Endoscope\_XX\_geometrical.xml}.
\item The Photometric Calibration parameters {\tt Endoscope\_XX\_photometrical.xml}.
\end{enumerate}
\subsubsection*{Geometric calibration}
The file {\tt Endoscope\_XX\_geometrical.xml} is the output calibration from Vicalib~\cite{Vicalib}. This XML file contains the intrinsic parameters of the camera ($f_x, f_y, C_x, C_y$,$k_1,k_2, k_3,k_4$) following the Vicalib output format.

\subsubsection*{Photometric calibration}

The photometric calibration file, \texttt{Endoscope\_XX\_photometrical.xml}, contains the calibrated parameters of the light source and the camera of the endoscope. An endoscope's \texttt{<rig>} may have one or more \texttt{<camera>} tags, associated with one or more \texttt{<light>} sources. Currently, only a single camera and a single virtual light are supported.

Each camera tag has a particular \texttt{<camera\_model>}. This model has a single parameter, the value of the gamma $\gamma$ response function in Eq.~\eqref{eq:photometric-model}. Regarding the light source, the \texttt{<light\_model>} has four parameters:  the intensity value $\sigma_0$, the light spread factor $\mu$ and two vectors for the light centre $\mathbf{P}$ and the principal direction $\mathbf{D}$.

\subsection*{Simulated colon}
All the data related to the simulated colon is included in the directory {\tt Simulated Sequences}. There is a directory {\tt seq\_X} ({\tt X} is the sequence number) for each sequence obtained from the simulation. The directory contains:

\begin{enumerate}
    \item The directory {\tt rgb} with the RGB images of the sequence in {\tt png} format.
    \item The directory {\tt depth} with the depth images for each RGB image of the sequence stored in {\tt exr} format.
    \item A file {\tt rgb.txt} with a list of file names of all RGB images of the sequence.
    \item A file {\tt depth.txt} with a list of the file names of all depth images of the sequence.
    \item A file {\tt trajectory.csv} containing the ground truth camera trajectory. 
    \item A file {\tt calibration.txt} containing the simulated camera calibration.
    \item A file {\tt info.txt} containing the deformations applied, its parameters and units.
\end{enumerate}

\subsection*{Meta-data}

This section contains the details and formats for each type of meta-data. The file {\tt DatasetSummary.xls} details the availability of the metadata in each sequence of the dataset.

\subsubsection*{Text footage}

 Two files: \texttt{Seq\_XXX.json} and \texttt{Seq\_XXX.srt} are included inside the root and \texttt{meta-data} folder. The \texttt{.json} file contains a structure with the timestamp and the associated text. The text footage is also included in \texttt{.srt} format to ease the visualization synchronized with the video. The references to identify the tools used during the procedure are stored in the \texttt{meta-data} directory.
 
\subsubsection*{Anatomical regions} Table~\ref{tab:places} shows the detailed number of frames labelled for each region in each video. The dataset contains this information in a file named \texttt{Anatomical\_Regions\_XXX.txt} with the format \texttt{Frame\#\#\#;region label;} in each line.

\subsubsection*{Tool Segmentation} 
\begin{figure}[htb]
\centering
\includegraphics[width=0.8\textwidth]{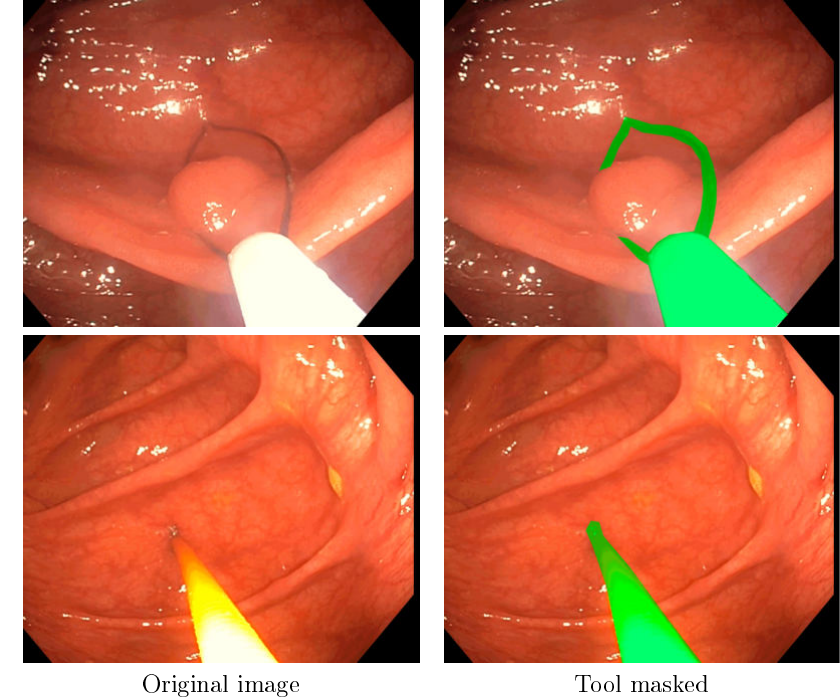}
\caption{Examples for the tool segmentation mask in Seq\_009}
\label{im:tool_seg}
\end{figure}

There are 4086 frames with tools segmented across four different colonoscopies as detailed in Table.~\ref{tab:seg}. The segmentations for each video can be found in file \texttt{tool\_segmentation\_XXX.xml}. This file contains, for each segmented frame, the id of the frame and a list of 2D point coordinates that define the tool segmentation as a binary polygon. The segmentation was done using a proprietary Odin CAT tool\cite{odinCat}. Some examples can be seen in Figure~\ref{im:tool_seg}.

\subsubsection*{COLMAP 3D reconstruction} 
\begin{figure}[htb]
\centering
\includegraphics[width=0.80\textwidth]{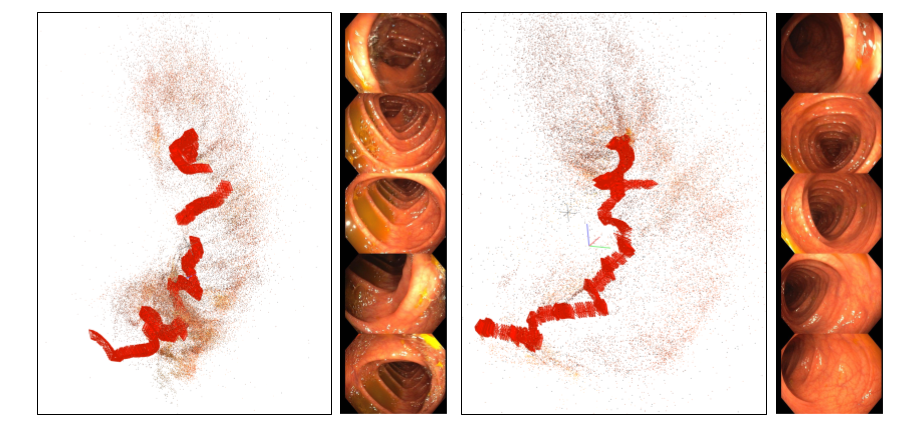}
\caption{Two clusters from the COLMAP reconstruction after processing Seq\_001. For each cluster, it is shown a 3D view of the frames' poses and colon map points and five RGB images as summary of the cluster frames.}
\label{fig:colmap}
\end{figure}

Table~\ref{tab:colmap} summarizes the reconstruction results for the Endomapper sequences. The reconstructions are stored following the text format of COLMAP (\url{https://colmap.github.io/format.html}). We provide text files showing the images contained in each cluster reconstructed by COLMAP, as well as the images that COLMAP considered covisible, i.e. images that have at least one 3D point in common. Figure~\ref{fig:colmap} shows two examples of these reconstructed clusters. 

\subsubsection*{Same patient recordings}
A file {\tt SamePatient.json} is stored in the root folder containing which sequences are from the same patient and the time that separates both sequences.

\begin{table}[htb]
\centering
\footnotesize
\begin{tabular}{|c|r|r|r|r|r|r|r|r|r|r|r|}
\hline
\textbf{Sections} & \textbf{Total Frames} & \textbf{rectum} & \textbf{sigmoid} & \textbf{descending} & \textbf{esplenic} & \textbf{transverse} & \textbf{hepatic} & \textbf{ascending} & \textbf{ileocecal} & \textbf{ileum} & \textbf{cecum} \\ \hline
\textbf{Seq\_003} & 78439 & 1519 & 2840 & 7640 & 320 & 3440 & 1840 & 58720 & 320 & 1800 & 0 \\ \hline
\textbf{Seq\_011} & 25840 & 1480 & 11920 & 4440 & 2200 & 3320 & 2480 & 0 & 0 & 0 & 0 \\ \hline
\textbf{Seq\_013} & 33360 & 360 & 2720 & 3600 & 2200 & 4200 & 2360 & 13400 & 1520 & 0 & 3000 \\ \hline
\textbf{Seq\_093} & 78480 & 920 & 14040 & 18960 & 6400 & 4640 & 2160 & 7200 & 2640 & 0 & 21520 \\ \hline
\textbf{Seq\_094} & 52560 & 400 & 21360 & 8400 & 1000 & 6000 & 760 & 8800 & 2400 & 0 & 3440 \\ \hline
\textbf{Total} & 268679 & 4679 & 52880 & 43040 & 12120 & 21600 & 9600 & 88120 & 6880 & 1800 & 27960 \\ \hline
\end{tabular}
\caption{\label{tab:places}Summary of the anatomical sections per video and label.}
\end{table}

\begin{table}[h]
\begin{minipage}[b]{85mm}
\centering
\begin{tabular}{|c|r|}
\hline
\multirow{2}{*}{Sequence} & Frames \\ 
 & segmented \\
\hline
Seq\_003 & 3168 \\ \hline
Seq\_013 & 254 \\ \hline
Seq\_093 & 435 \\ \hline
Seq\_094 & 229 \\ \hline

\end{tabular}
    \caption{ \label{tab:seg}Summary of the frames with tool segmentation.} 
\end{minipage}
\begin{minipage}[b]{85mm}
\centering
\begin{tabular}{|c|r|r|r|}
\hline
\multirow{2}{*}{Sequence} & Total & Frames & Clusters \\ 
 & frames & reconstructed & obtained \\
\hline
Seq\_001 & 14824 &  5809 ($39,18 \%$) & 50 \\ \hline
Seq\_002 & 23375 &  8133 ($34,79 \%$) & 50 \\ \hline
\end{tabular}
    \caption{\label{tab:colmap}Summary of COLMAP 3D reconstruction. } 
\end{minipage}

 \end{table}

\section*{Technical validation}
Firstly, we detail an error analysis of the calibration and a comparison of the calibration parameters among endoscopes. Secondly, we test state-of-the-art SfM and VSLAM algorithms on typical colonoscopy sequences. Finally, the anatomical region and tool segmentation labels are validated on state-of-the-art recognition algorithms. All the code used in this section is publicly available in the Endomapper repositories~\url{https://github.com/Endomapper}.

\subsection*{Calibration validation}
\paragraph{Geometric calibration}\label{geocalibration}

 The software used to compute the geometric calibration and to obtain the validation and comparisons shown in this section is available in the repository \textit{EM\_Dataset-GeometricCalibration} (see Section \textit{Code availability},~\url{https://github.com/endomapper/EM_Dataset-GeometricCalibration}). The geometric calibration was computed from the calibration videos using the Vicalib~\cite{Vicalib} tool, tuning the parameters for each endoscope calibration separately. The parameters are detailed in the repository.

\begin{figure*}[!t]
\centering
\includegraphics[width=\textwidth]{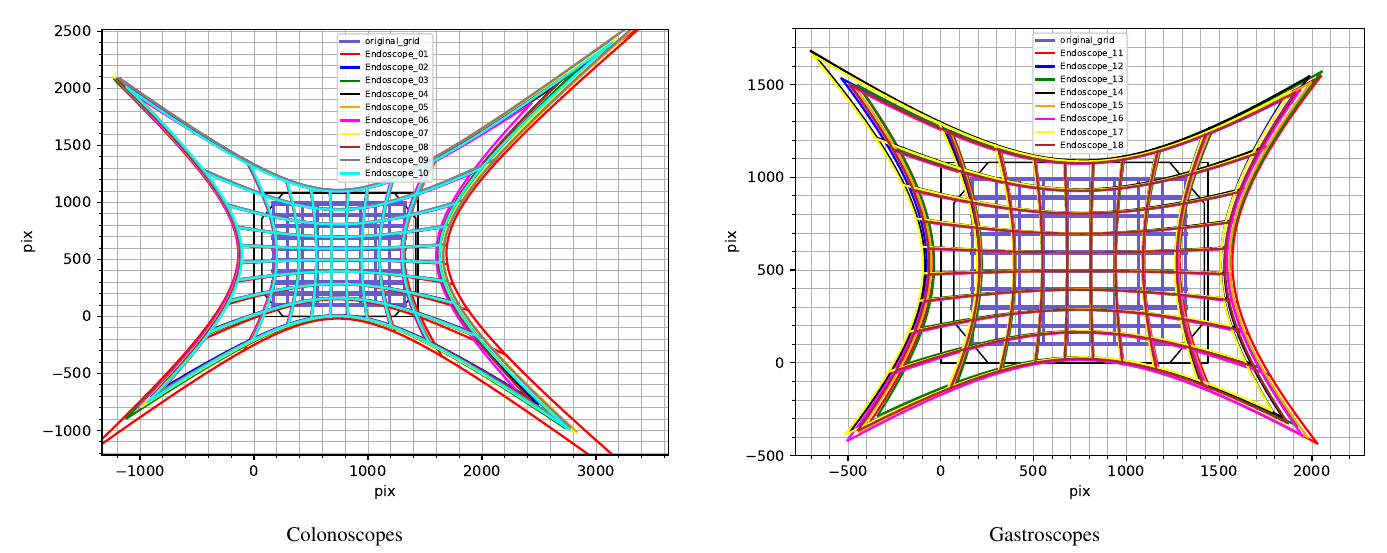}
\caption{Comparison of how a regular pixel grid is undistorted by the calibration of each endoscope. Colonoscopes and gastroscopes are separated for easier visualization.}
\label{fig:grid}
\end{figure*}

To compare the different calibrations visually, we have undistorted a grid using each calibration. Figure~\ref{fig:grid} shows the differences between each endoscope, the 10 colonoscopes and 8 gastroscopes correspondingly. The results show that the calibrations are equivalent around the center of the images and differences between them are significant only in the image borders. 

\begin{figure*}[!t]
\centering
\includegraphics[width=\textwidth]{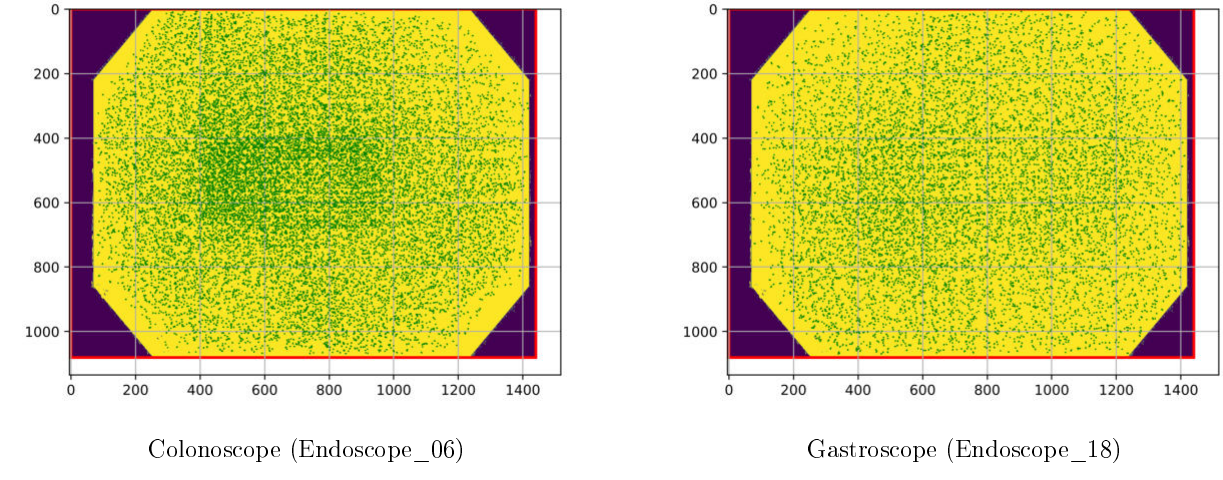} 
\caption{Distributions of error in the images in prototype calibrations. The line representing the error is not magnified, observe that most of them appear as points as errors are mostly smaller than one pixel.}
\label{Error_dist}

\end{figure*}

To further analyze the calibration results, we analyzed the reprojection error. For all the calibrations, the RMSE is between 0.3 and 0.4 pixels. We have selected two endoscopes, one colonoscope (Endoscope\_06) and one gastroscope (Endoscope\_18), as prototypes. Figure~\ref{Error_dist} displays the inliers reprojection error distribution for the selected endoscopes. Here we can see that the error of the inliers is uniformly distributed around the image, and that only at the image boundaries there are fewer measurements. The calibrations are then expected to be very accurate in general, being the most inaccurate areas the ones closer to the borders. 

\begin{figure*}[!t]
\centering

\includegraphics[width=\textwidth]{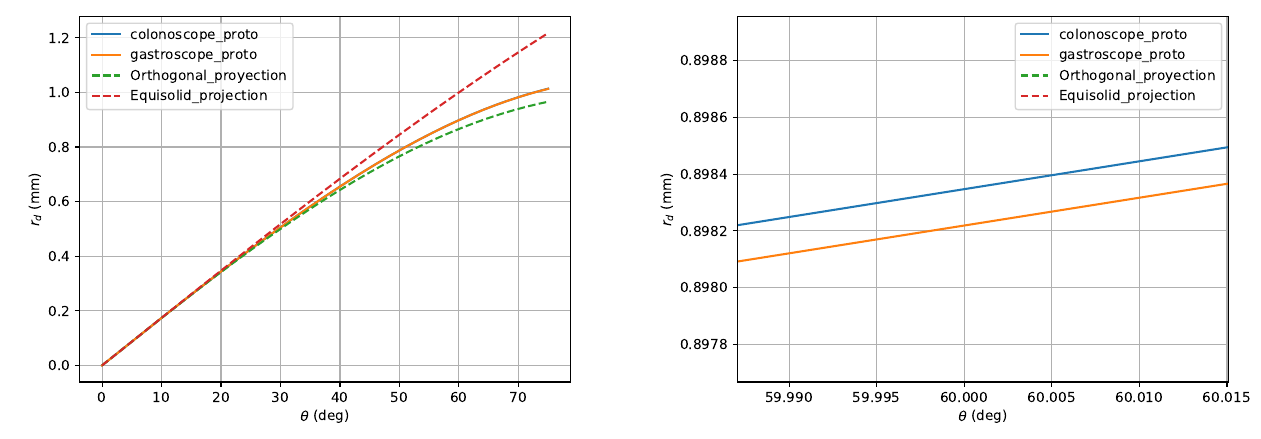}
 
\caption{Relation between the incoming ray angle $\theta$ with the distorted radius $r_d$. The dotted curves represent the ideal orthogonal and equisolid projection models. The right image is a zoom of the curves to show the small differences between the colonoscope and the gastroscope. }
\label{im:Calib_graph}
\end{figure*}

\begin{figure*}[!t]
\centering
\includegraphics[width=\textwidth]{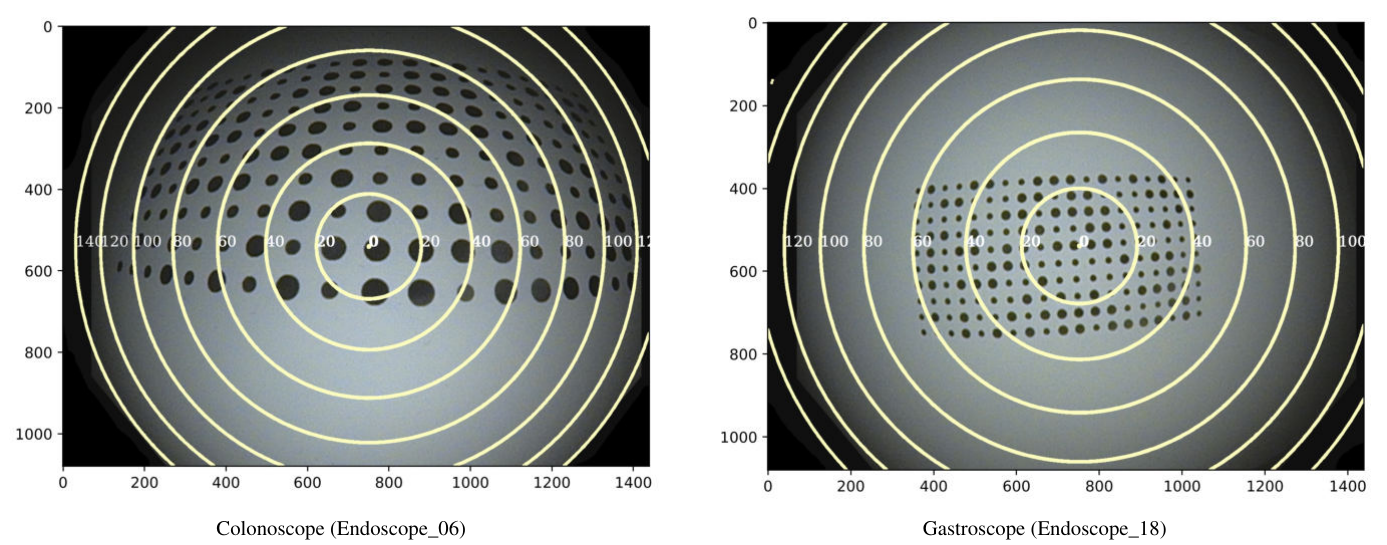} 
\caption{View angles plotted on top of calibration images from each prototype endoscope. The iso-lines are plotted in $20 ^{\circ}$ intervals.}
\label{im:view_angle}
\end{figure*}

The projection function that relates the incoming ray angle $\theta$ with the distorted radius $r_d$ is plotted in Figure~\ref{im:Calib_graph}. Here we can see that both types of endoscopes are almost equal and very close to an orthogonal projection fisheye lens \cite{kannala2006generic}. Finally, in Figure~\ref{im:view_angle} we show the view angle of both prototype endoscopes. There we can see that the gastroscope has a lower view angle than the colonoscope on the edges, which is why each type of endoscope needs to be calibrated separately.

With this analysis we conclude that the individual calibrations obtained from the videos are accurate (RMSE is low, covers most image and is consistent with all endoscopes). We believe that accurate calibration information boosts the performance of the geometric methods.

\begin{table}
\centering
\begin{tabular}{|c|c|c|c|c|c|c|c|c|c|}
\hline
\textbf{Colonoscope} & 02 & 03 & 04 & 05 & 06 & 07 & 08 & 09 & Mean\\
\hline
\textbf{RMSE} & 3.10 & 3.09 & 2.46 & 2.47 & 3.08 & 3.74 & 2.70 & 2.90 & 2.94  \\ 
\hline
\hline
\textbf{Gastroscope} & 11 & 12 & 13 & 14 & 15 & 16 & 17 & 18 & Mean\\
\hline
\textbf{RMSE} & 3.54 & 3.07 & 3.50 & 3.55 & 3.15 & 3.36 & 3.60 & 2.96 & 3.34  \\ 
\hline
\end{tabular}
    \caption{\label{tab:photo}Summary of photometric calibration errors.} 
\end{table}

\paragraph*{Photometric calibration}
The software used to compute the photometric calibration and to obtain the validation shown in this section is available in the repository \textit{EM\_Dataset-PhotometricCalibration} (see Section \textit{Code availability},~\url{https://github.com/endomapper/EM_Dataset-PhotometricCalibration}). The photometric calibration was computed from the calibration videos for each endoscope separately.

We selected 38 frames per sequence, looking for a variety of distances from the camera to the calibration pattern. On each frame, we consider a $120^\circ$ field of view. The centre of the virtual light converges about 4~mm behind the tip of the endoscope, thus being able to model all real lights with a single beam. The gamma value is experimentally fixed to $\gamma = 2.2$, which is also a broadly used value. The endoscope applies a continuous gain control, progressively increasing or decreasing the gain. Relative auto-gain is estimated with respect to the first image of the sequence. Consequently, the $\sigma_0$ value is unobservable and it is fixed to one.

The resulting models are validated in a different set of images of the Vicalib pattern. The photometric errors in Table~\ref{tab:photo} show the validation results of eight colonoscopes and eight gastroscopes. In the colonoscopes, the calibration is able to estimate the pixel intensities of the images with an RMSE of 2.9 grey levels. In gastroscopes, lights are not symmetrically placed on the tip of the endoscope. Consequently, RMSE increases slightly, up to 3.3 grey levels.

\subsection*{SfM/SLAM validation}

\paragraph*{COLMAP validation}
COLMAP is able to estimate sparse reconstructions for different sections along a sequence, see some examples in Figure \ref{fig:colmap}. As it can be seen, the 3D point cloud and the camera trajectory look reasonable, showing a tubular shape with cameras traversing it. The covisibility information is always accurate, as the geometrical checks in COLMAP avoid frames that do not observe the same place to be incorrectly reconstructed in the same model.

\begin{figure}[!h]
\centering
    \includegraphics[width=0.7\textwidth]{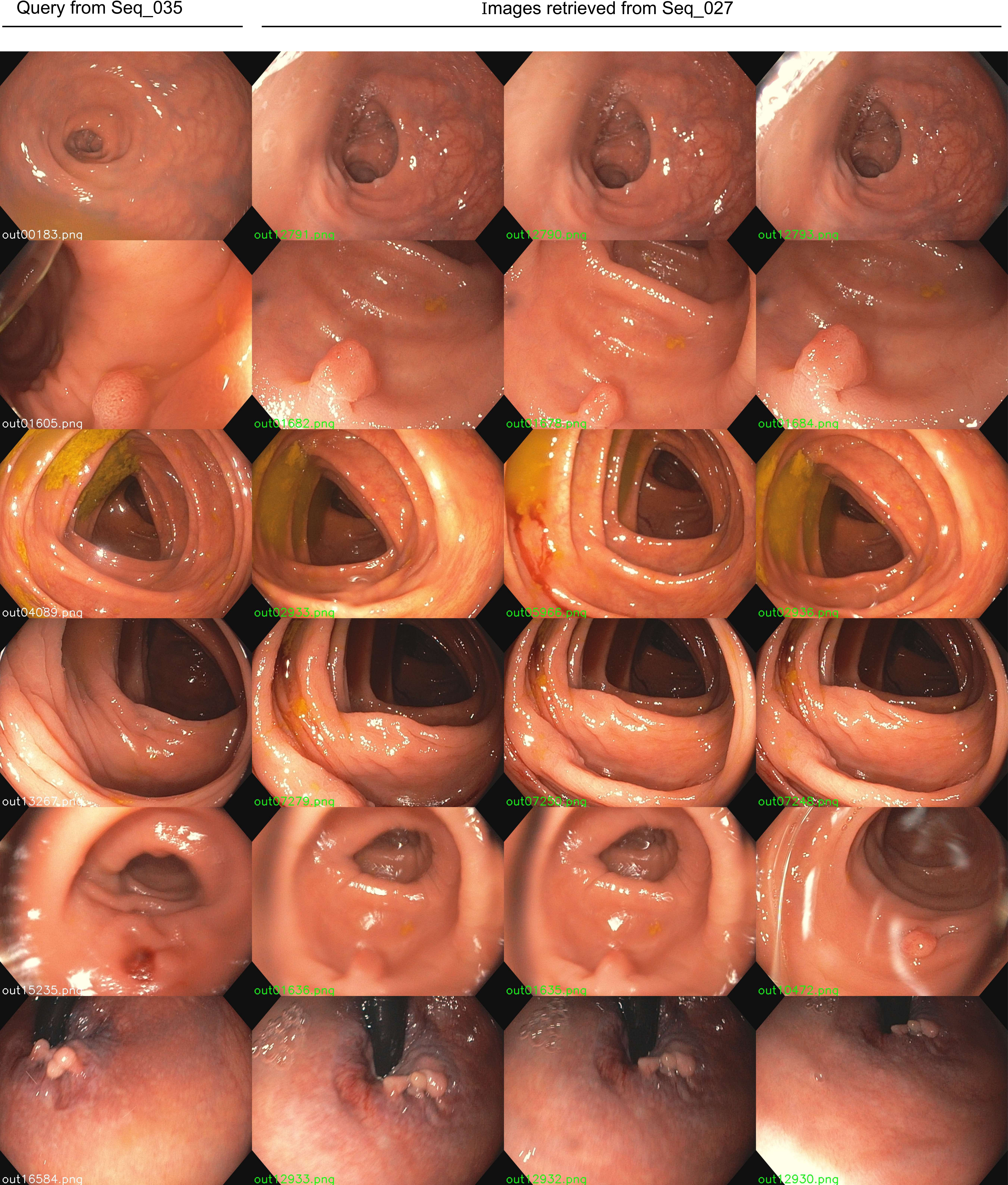} 
    \caption{Examples of successful retrieval from different sequences of the same patient. The left column contains the queries from the current sequence (Seq\_035) while the rest of the columns are the first three retrieved images from the previous sequence (Seq\_027).}
    \label{fig:same_patient}
\end{figure}

Covisibility information was exploited in \cite{morlana2021self}, where a CNN was trained with COLMAP reconstructions from our sequences for the task of image retrieval. The system \cite{morlana2021self} is able to recognize frames observing the same place in the colon, even when the frames come from different colonoscopies of the same patient. In Figure \ref{fig:same_patient}, we can see some examples of successfully retrieved between two colonoscopies of the same patient performed within two weeks of each other. 

The 3D reconstructions look reasonably accurate and can be of great help as a weak supervision for training depth, camera pose or image retrieval networks. The software used to compute the reconstruction shown in this section is available in the repository \textit{EM\_Dataset-ColmapValidation} (see Section \textit{Code availability},~\url{https://github.com/endomapper/EM_Dataset-ColmapValidation}).

\paragraph*{ORB-SLAM validation}

\begin{table}[htb]
\centering
\footnotesize
\begin{tabular}{|c|r|r|r|}
\hline
\textbf{} & \textbf{\# Keyframes} & \textbf{\# Map points} & \textbf{\# Frames}  \\ \hline
\textbf{Mean} & 22 & 1444 & 91  \\ \hline
\textbf{Median} & 18 & 1340 & 57 \\ \hline
\textbf{Min} & 11 & 398 & 16  \\ \hline
\textbf{Max} & 60 & 4145 & 643 \\ \hline
\end{tabular}
\caption{ Summary of the size of the 133 sub-maps obtained after processing Seq\_015}
\label{tab:orb_maps}
\end{table}

ORB-SLAM3\cite{campos2021orb} is the reference system for sparse real-time visual SLAM. We have used it to process the whole Seq\_015 video, to build the map and estimate the endoscope pose. To achieve a real-time performance the image size is reduced from $1440\times1080$ to $720\times540$ and one out of every two frames are skipped. To address non-rigidity, the reprojection error acceptance threshold has been increased by a $\times 2$ factor with respect to its default value for rigid scenes, which helps in preventing tracking losses. The Kannala-Brandt camera model has proven to be essential to extract and triangulate features close to the borders of the image, where distortion is significant. As argued, an accurate calibration enables the use of the whole image for geometric computation, boosting accuracy and robustness.

\begin{figure}[htb]
\centering
    \includegraphics[width=0.7\textwidth]{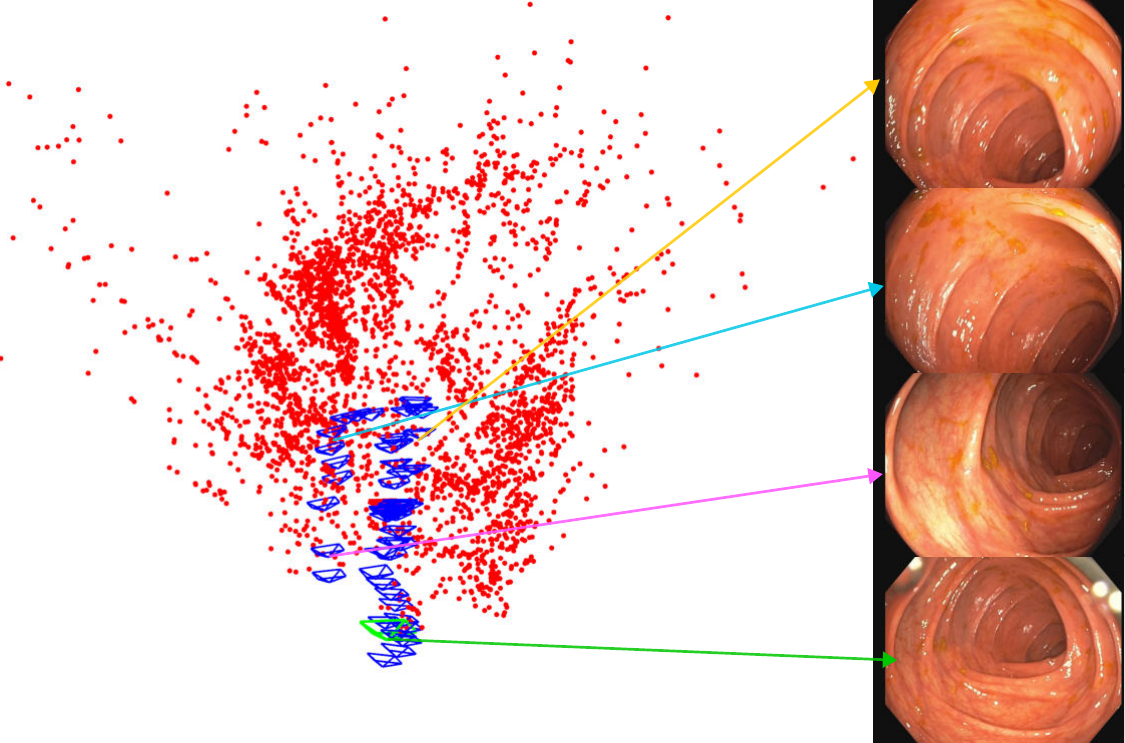} 
    \caption{ORB-SLAM3 sub-map in Seq\_015 between frames 54420 and 55170. The camera undergoes a forward-backwards motion. Right, 3D map in top view, keyframes in blue, map points in red. Left images corresponding to 4 keyframes spread over the trajectory.}
    \label{fig:orb_slam3_track}
\end{figure}

ORB-SLAM3 has been able to estimate 133 sub-maps of small size (see Table \ref{tab:orb_maps} for a summary of the map specifications). The camera has been localized successfully with respect to a map in 25\% of the frames. The time between video frames is 40 ms, and
ORB-SLAM3 is able to run in real time, with an average tracking time of 23 ms and maximum of 37 ms. Figure \ref{fig:orb_slam3_track} shows a sub-map where the camera undergoes a forward-backward motion The map contains 54 keyframes, 3682 points and 349 frames.

From this analysis we conclude that our EndoMapper dataset offers the challenges of real endoscopy exploration such as scene deformation, multiple occlusion, changes in lighting, and clutter due to cleaning water or tools that eventually result in frequent tracking losses. Classical discrete feature VSLAM methods like ORB-SLAM3 can run on these videos in real-time, localizing the camera in 25\% of the frames. However, the scene model is fragmented in a myriad of very small rigid sub-maps. The clear challenge is multiple mapping techniques operating in Endoscopy able to merge sub-maps with common areas.

 The software and the detailed tuning used to compute the ORB-SLAM3 reconstructions shown in this section is available in the repository \textit{EM\_Dataset-ORBSLAM3Validation} (see Section \textit{Code availability},~\url{https://github.com/endomapper/EM_Dataset-ORBSLAM3Validation}).

\subsection*{Anatomical region validation}
 The anatomical region labels have been validated by fine-tuning different models for anatomical region recognition. The software used to evaluate the anatomical regions recognition is available in the repository \textit{EM\_Dataset-AnatomicalRegions} (see Section \textit{Code availability},~\url{https://github.com/endomapper/EM_Dataset-AnatomicalRegions}). Following works in medical image~\cite{thanh2020polyp} and datasets~\cite{borgli2020hyperkvasir}, we fine-tuned four different CNNs that are known to perform well on medical image classification: EfficientNet V2~\cite{tan2021efficientnetv2}, MobileNetv2~\cite{sandler2018mobilenetv2}, DenseNet~\cite{huang2017densely} and ResnetV2~\cite{he2016identity}. With MobileNet and EfficientNet, we look for a model requiring low computational resources. DenseNet and ResNet were chosen for its performance in image classification. For the fine-tuning, we train the models during 100 epochs following the learning rate decay in Thanh et al.~\cite{tan2021efficientnetv2} and use 4 sequences (Seq\_003,
 Seq\_011, Seq\_013, Seq\_093) for training and Seq\_094 for testing. Seq\_094 was chosen as the test sequence because it has the best balance between classes. The metrics used to evaluate this experiment were Top-1 and Top-3 accuracy, defined as the accuracy for which the true class matches the most probable prediction and any of the 3 most probable predictions, respectively. Table~\ref{Anatomical_res} presents the results for the  anatomical region recognition.
 
 ResNet is able to perform better in Top-1 accuracy and similarly in Top-3. This shows that this model is the best overall. It is also interesting to note that MobileNet is able to obtain a close performance with a smaller computational footprint, being interesting for real-time systems. It is also worth remarking that Top-1 accuracy is low in comparison to other computer vision tasks, suggesting that anatomical region classification is a challenging research problem. Indeed, the differences between sections are very subtle, even for the trained eye. In any case, the results on Top-3 accuracy are promising and show that the EndoMapper data is a relevant tool to advance the performance in this problem.

\begin{table}[htb]
\centering
\footnotesize
\begin{tabular}{|l|r|r|}
\hline
\textbf{Model} & \textbf{Top-1 Accuracy (\%)} & \textbf{Top-3 Accuracy (\%)} \\ \hline
\textit{EfficientNet V2-M} & 17.61 & 48.00 \\ \hline
\textit{MobileNet V2} & 20.40 & 60.17 \\ \hline
\textit{DenseNet-161} & 24.87 & \textbf{66.77} \\ \hline
\textit{ResNetV2-101} & \textbf{26.08} & 66.09 \\ \hline
\end{tabular}
\caption{Top-1 and Top-3 accuracy of the anatomical region recognition models. All models were fine-tuned in 4 sequences(Seq\_003,Seq\_011,Seq\_013,Seq\_093) on top of ImageNet pre-trained weights, and tested on Seq\_094.}
\label{Anatomical_res}

\end{table}

\subsection*{Tools segmentation validation}

\begin{figure}[!bt]
    \centering
    \includegraphics[width=\textwidth]{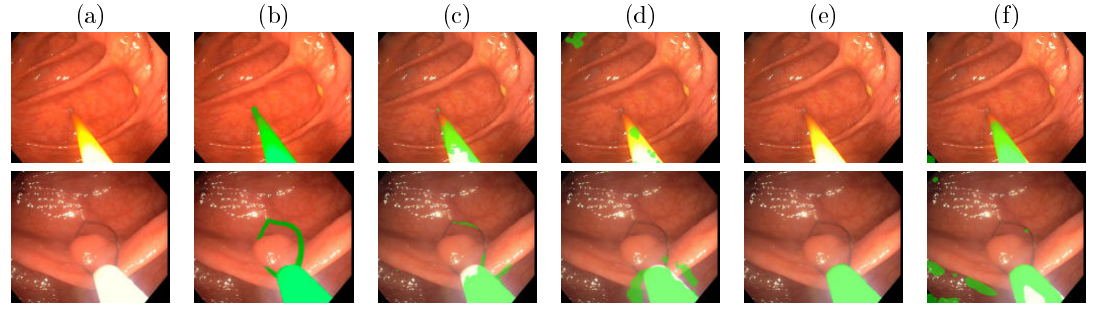}
    \caption{\textbf{Binary Segmentation} examples from Endomapper dataset using different approaches fine-tuned on Endomapper dataset: (a) Original Image (b) Ground-truth manual segmentation (c) MiniNet (d) UNet (e) LinkNet (f) MF-TAPNet}
    \label{fig:binary_mininet}
\end{figure}

The tool segmentation labels have been validated by training and testing several models for binary tool segmentation as proposed in Tomasini et al.\cite{tomasini2022efficient}. This work compares the performance of various models on three different datasets, including the EndoMapper dataset labels. All the models were trained from scratch on EndoVis17 dataset and fine-tuned on Kvasir-Inst and EndoMapper. The performance results obtained can be seen in Table~\ref{tab:binary-res}. Examples of binary segmentation of images from the Endomapper dataset obtained using the different models can be seen in Figure~\ref{fig:binary_mininet}.

The lower mIoU of all models on our EndoMapper dataset compared to that of other datasets highlights the challenge of the EndoMapper tool segmentation labels. It is interesting to note that the efficient MiniNetV2 reaches similar performance to state-of-the-art models on all datasets while requiring less memory and inference time. The software used to evaluate the tool segmentation models is available in the repository \textit{EM\_Dataset-ToolSegmentation} (see Section \textit{Code availability},~\url{https://github.com/endomapper/EM_Dataset-ToolSegmentation}).

\begin{table}[!htb]
  \small
  \centering
  \begin{tabular}{|@{}l||r|r|r||r|r|}
  \hline
  \bfseries  & \multicolumn{3}{c||}{Datasets} & \multicolumn{2}{c|}{Computational cost} \\
  \bfseries Models & \bfseries EndoVis17 & \bfseries Kvasir-Inst. & \bfseries Endomapper & \bfseries Params (M)$^+$ & \bfseries Time (ms)$^{++}$ \\

  \hline
  U-Net\cite{ronneberger2015u} 
  & 75.44 & 85.78 & 55.63 & 7.85 & 54
  \\
  \hline
  TernausNet\cite{iglovikov2018ternausnet} 
  & 83.60 & N.A. & N.A. & 36.92 & 119
  \\
  \hline
  LinkNet\cite{chaurasia2017linknet} 
  
  & 82.36 & 87.75 & 60.54 & 21.79 & 34
  \\
  \hline
  MF-TAPNet\cite{jin2019incorporating} 
  & 87.56 & 86.81 & 66.87
  & 37.73 & 155
  \\
  \hline
  MiniNet v2\cite{alonso2020mininet} 
  & 87.16 & 85.13 & 66.65 & 0.52 & 26 \\
  \hline 
  \multicolumn{6}{l}{\footnotesize $^+$: \textit{Params}: Memory required by the model (M = millions of parameters).}\\
  \multicolumn{6}{l}{\footnotesize $^{++}$: \textit{Time}: Average inference time for 1 image on GPU RTX2080.}
  \end{tabular}
   \caption{\textbf{Binary segmentation} results (\textbf{mIoU}) for models pre-trained on EndoVis17 and fine-tuned for each of the target datasets (Kvasir and Endomapper). N.A.: Not available due to computational resource limitations.}%

  \label{tab:binary-res}
\end{table}

\section*{Usage notes}
\label{sec_usage_notes}
The dataset is available on the Synapse platform~\cite{EndomapperDataset}. The dataset can be publicly accessed under the following
conditions: 
1) Limited to research on how to obtain relevant medical information from images or video. 2) Redistribution of the data is not allowed.
3) Requires a Statement of Intended Use, which includes a description of how you intend to use this data.
4) You further agree to cite the DOI of the collection and the publication in any publication resulting from this content as follows:
a) Azagra, P. et al. Endomapper dataset of complete calibrated endoscopy procedures. \url{https://doi.org/10.7303/syn26707219} (2022). Synapse.
b) Azagra, P. et al. EndoMapper dataset of complete calibrated endoscopy procedures. Scientific Data. 
5) Images of the collection can be included in the scientific citing publications.
6) Video segments can be used to produce multimedia material in the citing scientific publications.

\clearpage

\section*{Code availability}

The dataset can be used without any further code. All the code used for the calibration, simulated sequences generation and technical validation are publicly available as repositories at ~\url{https://github.com/Endomapper}.  The instructions for installing and using them are available in Synapse~\url{https://www.synapse.org/#!Synapse:syn52137895}.

\bibliography{Main_final}

\section*{Acknowledgements} 

This work was supported by EU-H2020 grant 863146 ENDOMAPPER, Spanish Government grants PI20/01514 and FPU20/06782, and Aragón Government grant T45 23R

\section*{Author contributions statement}

P.A. coordinated the dataset acquisition and lead the organization.
J.M.M.M. originated the concept of dataset.
P.A., L.R., J.C., J.D.T., A.C.M. and J.M.M.M. designed the dataset details.
C.S., A.F., and A.L. performed the endoscopies, provided medical explanations and anatomical labels.
P.A., L.R., C.O. and J.M.M.M. designed and operated the data acquisition system and created the database.
P.A., V.M.B., J.D.T. and J.M.M.M. performed endoscope's calibration.
J.J.G.R. and J.D.T. provided colon simulations.
O.L.B., J.M. and J.M.M.M. provided COLMAP reconstructions.
C.T., L.R. and A.C.M provided tool segmentation.
P.A. and J.L. provided anatomical landmark annotations.
P.A., L.R., O.L.B., C.T., J.M., D.R., V.M.B., J.J.G.R., R.E., J.C., J.D.T., A.C.M. and J.M.M.M performed the analysis and technical validation. 
P.A., L.R., J.M., V.M.B., J.J.G.R., J.D.T. and J.M.M.M. created and edited the manuscript. 
All authors reviewed the manuscript.

\section*{Competing interests} 

The authors declare no competing interests.

\end{document}